\documentclass[12pt]{article}

\renewcommand{\theequation}{\arabic{section}.\arabic{equation}}

\def\hybrid{\topmargin -20pt    \oddsidemargin 0pt
        \headheight 0pt \headsep 0pt

        \textwidth 6.25in       
        \textheight 9.5in       
        \marginparwidth .875in
        \parskip 5pt plus 1pt   \jot = 1.5ex}

\hybrid

\catcode`\@=11

\def\marginnote#1{}

\newcommand{\vdd}[2]{\frac{\delta #1}{\delta #2}}
\newcommand{\eqref}[1]{(\ref #1)}

\def\d{\partial}

\newcommand{\vddr}[2]{\frac{\delta^R #1}{\delta #2}}
\newcommand{\vddl}[2]{{\frac{\delta^L #1}{\delta #2}}}

\def\half{{\frac{1}{2}}}
\newcommand{\p}[1]{{\rm p}(#1)}

\def\theequation{\thesection.\arabic{equation}}

\def\be{\begin{eqnarray}}
\def\ee{\end{eqnarray}}
\def\beann{\begin{eqnarray*}}
\def\eeann{\end{eqnarray*}}
\def\beq{\begin{equation}}
\def\eeq{\end{equation}}
\def\ba{\begin{array}}
\def\ea{\end{array}}
\def\ben{\begin{enumerate}}
\def\een{\end{enumerate}}
\def\bea{\begin{eqnarray}}
\def\eea{\end{eqnarray}}
\def\beann{\begin{eqnarray*}}
\def\eeann{\end{eqnarray*}}
\def\beq{\begin{equation}}
\def\eeq{\end{equation}}
\def\ba{\begin{array}}
\def\ea{\end{array}}
\def\ben{\begin{enumerate}}
\def\een{\end{enumerate}}

\def\5{\bar }
\def\6{\partial }
\def\7{\hat }
\def\4{\tilde }

\def\cF{{\cal F}}

\def\s0#1#2{\mbox{\small{$\frac{#1}{#2}$}}}

\def\f#1#2#3{{f_{#1#2}}^{#3}}

\def\G{\Gamma}
\def\D{\Delta}

\newcount\hour
\newcount\minute
\newtoks\amorpm
\hour=\time\divide\hour by60
\minute=\time{\multiply\hour by60 \global\advance\minute by-\hour}
\edef\standardtime{{\ifnum\hour<12 \global\amorpm={am}%
        \else\global\amorpm={pm}\advance\hour by-12 \fi
        \ifnum\hour=0 \hour=12 \fi
        \number\hour:\ifnum\minute<10 0\fi\number\minute\the\amorpm}}
\edef\militarytime{\number\hour:\ifnum\minute<10 0\fi\number\minute}

\def\draftlabel#1{{\@bsphack\if@filesw {\let\thepage\relax
   \xdef\@gtempa{\write\@auxout{\string
      \newlabel{#1}{{\@currentlabel}{\thepage}}}}}\@gtempa
   \if@nobreak \ifvmode\nobreak\fi\fi\fi\@esphack}
        \gdef\@eqnlabel{#1}}
\def\@eqnlabel{}
\def\@vacuum{}
\def\draftmarginnote#1{\marginpar{\raggedright\scriptsize\tt#1}}

\def\draft{\oddsidemargin -.5truein
        \def\@oddfoot{\sl preliminary draft \hfil
        \rm\thepage\hfil\sl\today\quad\militarytime}
        \let\@evenfoot\@oddfoot \overfullrule 3pt
        \let\label=\draftlabel
        \let\marginnote=\draftmarginnote
   \def\@eqnnum{(\theequation)\rlap{\kern\marginparsep\tt\@eqnlabel}%
\global\let\@eqnlabel\@vacuum}  }

\def\preprint{\twocolumn\sloppy\flushbottom\parindent 2em
        \leftmargini 2em\leftmarginv .5em\leftmarginvi .5em
        \oddsidemargin -.5in    \evensidemargin -.5in
        \columnsep .4in \footheight 0pt
        \textwidth 10.in        \topmargin  -.4in
        \headheight 12pt \topskip .4in
        \textheight 6.9in \footskip 0pt
        \def\@oddhead{\thepage\hfil\addtocounter{page}{1}\thepage}
        \let\@evenhead\@oddhead \def\@oddfoot{} \def\@evenfoot{} }

\def\numberbysection{\@addtoreset{equation}{section}
        \def\theequation{\thesection.\arabic{equation}}}

\def\underline#1{\relax\ifmmode\@@underline#1\else
        $\@@underline{\hbox{#1}}$\relax\fi}

\def\titlepage{\@restonecolfalse\if@twocolumn\@restonecoltrue\onecolumn
     \else \newpage \fi \thispagestyle{empty}\c@page\z@
        \def\thefootnote{\fnsymbol{footnote}} }

\def\endtitlepage{\if@restonecol\twocolumn \else \newpage \fi
        \def\thefootnote{\arabic{footnote}}
        \setcounter{footnote}{0}}  

\catcode`@=12
\relax

\makeatletter
\newcounter{pubctr}
\def\publist{\@ifnextchar[{\@publist}{\@@publist}}
\def\@publist[#1]{\list
        {[\arabic{pubctr}]\hfill}{\settowidth\labelwidth{[999]}
        \leftmargin\labelwidth
        \advance\leftmargin\labelsep
        \@nmbrlisttrue\def\@listctr{pubctr}
        \setcounter{pubctr}{#1}\addtocounter{pubctr}{-1}}}
\def\@@publist{\list
        {[\arabic{pubctr}]\hfill}{\settowidth\labelwidth{[999]}
        \leftmargin\labelwidth
        \advance\leftmargin\labelsep
        \@nmbrlisttrue\def\@listctr{pubctr}}}
 \relax
\makeatother
\newskip\humongous \humongous=0pt plus 1000pt minus 1000pt

\newif\ifdtup

\relax

\def\d{\partial}

\def\sqr#1#2{{\vcenter{\vbox{\hrule height.#2pt\hbox{\vrule width.#2pt
height#1pt \kern#1pt \vrule width.#2pt}\hrule height.#2pt}}}}

\def\=d{\,{\buildrel\rm def\over =}\,}

\def\F{{\cal F}}

\def\i3p{\p32\int d^3p}

\def\As{A\hbox to 1pt{\hss /}}
\def\np4{\int d^4p_1\cdots d^4p_{n-1}\, }

\def\nx4{\int d^4x_1\ldots d^4x_n\, }

\def\kon#1#2{\vbox{\halign{##&&##\cr
\lower4pt\hbox{$\scriptscriptstyle\vert$}\hrulefill &
\hrulefill\lower4pt\hbox{$\scriptscriptstyle\vert$}\cr $#1$&
$#2$\cr}}}

\def\konv#1#2#3{\hbox{\vrule height12pt depth-1pt}
\vbox{\hrule height12pt width#1cm depth-11.6pt}
\hbox{\vrule height6.5pt depth-0.5pt}
\vbox{\hrule height11pt width#2cm depth-10.6pt\kern5pt
      \hrule height6.5pt width#2cm depth-6.1pt}
\hbox{\vrule height12pt depth-1pt}
\vbox{\hrule height6.5pt width#3cm depth-6.1pt}
\hbox{\vrule height6.5pt depth-0.5pt}}
\def\konu#1#2#3{\hbox{\vrule height12pt depth-1pt}
\vbox{\hrule height1pt width#1cm depth-0.6pt}
\hbox{\vrule height12pt depth-6.5pt}
\vbox{\hrule height6pt width#2cm depth-5.6pt\kern5pt
      \hrule height1pt width#2cm depth-0.6pt}
\hbox{\vrule height12pt depth-6.5pt}
\vbox{\hrule height1pt width#3cm depth-0.6pt}
\hbox{\vrule height12pt depth-1pt}}

\def\konw#1#2#3{\hbox{\vrule height12pt depth-1pt}
\vbox{\hrule height12pt width#1cm depth-11.6pt}
\hbox{\vrule height6.5pt depth-0.5pt}
\vbox{\hrule height12pt width#2cm depth-11.6pt \kern5pt
      \hrule height6.5pt width#2cm depth-6.1pt}
\hbox{\vrule height6.5pt depth-0.5pt}
\vbox{\hrule height12pt width#3cm depth-11.6pt}
\hbox{\vrule height12pt depth-1pt}}

\def\i{{\rm int}}

\def\e{{\rm ext}}

\def\r{{\rm ret}}
\def\a{{\rm av}}

\def\m3{{\mu_1\mu_2\mu_3}}

\def\p{{(+)}}

\def\be{\begin{equation}}       \def\eq{\begin{equation}}
\def\ee{\end{equation}}         \def\eqe{\end{equation}}

\def\bea{\begin{eqnarray}}      \def\eqa{\begin{eqnarray}}
\def\ena{\end{eqnarray}}        \def\eea{\end{eqnarray}}
                                \def\eqae{\end{eqnarray}}

\def\ba{\begin{array}}
\def\ea{\end{array}}
\def\unit{1 \hskip-.3em \raise2pt\hbox{$ \scriptstyle |$ } }

\def\a{\alpha}

\def\d{\delta}
\def\e{\epsilon}           
\def\f{\phi}               
\def\g{\gamma}

\def\i{\iota}

\def\l{\lambda}
\def\m{\mu}
\def\n{\nu}
  
\def\p{\pi}                
\def\r{\rho}                                     
\def\s{\sigma}                                   
\def\t{\tau}

\def\D{\Delta}
\def\F{\Phi}
\def\G{\Gamma}

\def\cj{{\cal J}}
\def\ck{{\cal K}}
\def\cl{{\cal L}}

\def\half{{1 \over 2}}

\def\bop#1{\setbox0=\hbox{$#1M$}\mkern1.5mu
        \vbox{\hrule height0pt depth.04\ht0
        \hbox{\vrule width.04\ht0 height.9\ht0 \kern.9\ht0
        \vrule width.04\ht0}\hrule height.04\ht0}\mkern1.5mu}
\def\pa{\partial}                              

\def\>{\rangle} 

\def\<{\langle} 
\def\Dsl{D \hskip-.6em \raise1pt\hbox{$ / $ } }

\def\sl#1{\rlap{\hbox{$\mskip 1 mu /$}}#1}
\def\leftrightarrowfill{$\mathsurround=0pt \mathord\leftarrow \mkern-6mu
       \cleaders\hbox{$\mkern-2mu \mathord- \mkern-2mu$}\hfill
       \mkern-6mu \mathord\rightarrow$}
\def\dvec#1{\vbox{\ialign{##\crcr
       \leftrightarrowfill\crcr\noalign{\kern-1pt\nointerlineskip}
       $\hfil\displaystyle{#1}\hfil$\crcr}}}          
\def\hook#1{{\vrule height#1pt width0.4pt depth0pt}}
\def\leftrighthookfill#1{$\mathsurround=0pt \mathord\hook#1
       \hrulefill\mathord\hook#1$}
\def\underhook#1{\vtop{\ialign{##\crcr                 
       $\hfil\displaystyle{#1}\hfil$\crcr
       \noalign{\kern-1pt\nointerlineskip\vskip2pt}
       \leftrighthookfill5\crcr}}}
\def\smallunderhook#1{\vtop{\ialign{##\crcr      
       $\hfil\scriptstyle{#1}\hfil$\crcr
       \noalign{\kern-1pt\nointerlineskip\vskip2pt}
       \leftrighthookfill3\crcr}}}

\def\sfrac#1#2{{\vphantom1\smash{\lower.5ex\hbox{\small$#1$}}\over
       \vphantom1\smash{\raise.4ex\hbox{\small$#2$}}}} 
\def\bfrac#1#2{{\vphantom1\smash{\lower.5ex\hbox{$#1$}}\over
       \vphantom1\smash{\raise.3ex\hbox{$#2$}}}}      
\def\afrac#1#2{{\vphantom1\smash{\lower.5ex\hbox{$#1$}}\over#2}}  
\def\on#1#2{{\buildrel{\mkern2.5mu#1\mkern-2.5mu}\over{#2}}}
\def\ddt#1{\on{\hbox{\LARGE .\kern-2pt.}}#1}             
\def\tdt#1{\on{\hbox{\LARGE .\kern-2pt.\kern-2pt.}}#1}   

\def\boxes#1{
       \newcount\num
       \num=1
       \newdimen\downsy
       \downsy=-1.5ex
       \mskip-2.8mu
       \bo
       \loop
       \ifnum\num<#1
       \llap{\raise\num\downsy\hbox{$\bo$}}
       \advance\num by1
       \repeat}
\def\boxup#1#2{\newcount\numup
       \numup=#1
       \advance\numup by-1
       \newdimen\upsy
       \upsy=.75ex
       \mskip2.8mu
       \raise\numup\upsy\hbox{$#2$}}

\newskip\humongous \humongous=0pt plus 1000pt minus 1000pt

\newif\ifdtup

\def\to{\rightarrow}

\def\1ov4{{1\over 4}}

\def\pa{\partial}

\def\ddt{\dot{\t}}

\def\pa{\partial}

\def\half{{1 \over 2}}

\def\a{\alpha}

\def\d{\delta}
\def\e{\epsilon}           
\def\f{\phi}               
\def\g{\gamma}

\def\l{\lambda}
\def\m{\mu}
\def\n{\nu}
  
\def\p{\pi}                
\def\r{\rho}                                     
\def\s{\sigma}                                   
\def\t{\tau}

\def\D{\Delta}
\def\F{\Phi}
\def\G{\Gamma}

\def\pa{\partial}

\def\cj{{\cal J}}
\def\ck{{\cal K}}
\def\cl{{\cal L}}

\def\be{\begin{equation}}       \def\eq{\begin{equation}}
\def\ee{\end{equation}}         \def\eqe{\end{equation}}

\def\bea{\begin{eqnarray}}      \def\eqa{\begin{eqnarray}}
\def\ena{\end{eqnarray}}        \def\eea{\end{eqnarray}}
                                \def\eqae{\end{eqnarray}}

\begin{document}

\thispagestyle{empty}

\setcounter{footnote}{0}

\begin{flushright}
CERN-TH/2003-127\\
SLAC-PUB-9929\\
ULB-TH-03/22\\
ITFA-2003-30\\
{\bf hep-th/0306127}
\end{flushright}

\vspace{1cm}

\begin{center}
{\Large \bf Comments on the Gauge Fixed BRST Cohomology \\
and the Quantum Noether Method}\footnote{Based on an invited lecture presented by T.H.
at the Hesselberg Meeting 2002
"Theory of Renormalization and Regularization", Hesselberg, Germany; a short version of the paper without background material will appear in Physics Letters B}
\end{center}

\vskip 0.5cm
\centerline{\bf Glenn Barnich\footnote{Research Associate of the National Fund for Scientific Research (Belgium)}}
\vskip 0,1cm
\centerline{\it Physique Th\'eorique et Math\'ematique, Universit\'e Libre de
Bruxelles}
\centerline{\it Campus Plaine C.P. 231, B-1050 Brussels, Belgium}
\centerline{\it gbarnich@ulb.ac.be}
  \vskip 0,5cm
\centerline{\bf Tobias Hurth\footnote{Heisenberg Fellow}}
\vskip 0,1cm
\centerline{\it CERN, Theory Division, CH-1211 Geneva 23, Switzerland}
\centerline{\it SLAC, Stanford University, Stanford, CA 94309, USA}
\centerline{\it tobias.hurth@cern.ch}
 \vskip 0,5cm
\centerline{\bf Kostas Skenderis}
  \vskip 0,1cm
\centerline{\it Institute for Theoretical Physics, University of Amsterdam}
\centerline{\it Valkenierstraat 65, 1018 XE Amsterdam, The Netherlands}
\centerline{\it skenderi@science.uva.nl}

\vspace{.5 cm}

{\bf Abstract.}
{\small
We discuss in detail the relation between the gauge fixed 
and gauge invariant BRST cohomology.
We showed previously that in certain gauges some 
cohomology classes of the gauge-fixed BRST differential 
do {\it not} correspond to gauge invariant observables.
We now show that in addition ``accidental'' conserved
currents may appear. These correspond one-to-one to observables 
that become trivial in this gauge. We explicitly show how the 
gauge-fixed BRST cohomology appears in the context of the 
Quantum Noether Method.
}

\vfill

\newpage

\section{Introduction}

\setcounter{equation}{0}

Noether's theorem in classical field
theory allows for an iterative construction of invariant
classical actions \cite{deser}.
Starting from a given Lagrangian that is invariant under some
symmetry, one can iteratively construct interactions
by adding extra terms to the action and to the transformation rules, so
 that the final action is invariant. In the simplest case
the starting Lagrangian is free.
In \cite{paper1, proc} a construction of theories with global (=rigid)
and/or local symmetries within the Bogoliubov--Shirkov--Epstein--Glaser
(BSEG) approach was presented.  This construction may be viewed as a
quantum version of the classical Noether method.

\medskip
In the BSEG approach,
the $S$-matrix is directly constructed in the Fock space of free
asymptotic fields as a formal power series. The coupling
constant is replaced by a tempered test function $g(x)$ (i.e. a smooth
function rapidly decreasing at infinity) which switches on the
interaction. Instead of obtaining the $S$-matrix by first computing
off-shell Green functions by means of Feynman diagrams and then
applying the LSZ formalism, the $S$-matrix is obtained by imposing
causality and Poincar\'{e} invariance. The method can be regarded as
an ``inverse'' of the cutting rules: one builds $n$-point functions
out of $m$-point functions ($m<n$) by suitably ``gluing'' them
together. The precise manner in which this is done is dictated by
causality and Poincar\'{e} invariance.  It can be shown that this process
uniquely fixes the $S$-matrix up to local terms (for a brief
introduction to the BSEG method see section 3 of \cite{paper1}; for
more detailed accounts the reader is referred to
\cite{BS,EG,EG2,stora,Scharf95}).

The BSEG approach should not be regarded as a special renormalization
scheme  but as a general framework in which the conditions posed by the
fundamental axioms of quantum field theory (QFT) on any
renormalization scheme are built in by construction. In this sense the
quantum Noether construction in \cite{paper1,proc} is independent from
the causal BSEG approach.  The proposed quantum Noether conditions
should hold in any other formalism.

\medskip

In the Lagrangian framework, powerful methods have been developed that
allow for a systematic cohomological analysis of local counterterms
and anomalies \cite{Batalin:1981jr}. The BRST-BV method provides an efficient
way of analysing Ward identities. In the BSEG operator
approach,  Ward identities play a central role: they provide the quantum
Noether conditions (QNC) \cite{paper1, proc}.  Consistency then
requires that in the (naive) adiabatic limit, $g(x) \rightarrow 1$,
the constraints imposed by the QNC be
equivalent to those obtained via the antifield
formalism. This issue was analysed in \cite{Barnich:1999cy}. We will
review and further elaborate on these results in this contribution.
In particular, we analyze in detail the issue of the coboundary
condition which we only touched upon in \cite{Barnich:1999cy}.

\medskip

The paper is organized as follows. In sections 2 and 3 the
antifield formalism is discussed.  Section 4 is devoted to the
discussion of the relation between the gauge fixed and gauge invariant
BRST cohomology.  In section 5 we present an example that illustrates
aspects of the analysis in section 4.
In section 6 we recall the quantum Noether
method and in section 7 we use the results of section 4 
to analyse the general
solution of the quantum Noether conditions.  The appendix contains a
brief review of the off-shell procedure used in BSEG computations.

\section{Renormalization with Antifields}

The problem of perturbative quantization of generic gauge theories in
the Lagrangian framework consists of renormalizing the theory while
preserving an appropriate form of gauge invariance at the quantum
level.  We briefly summarize the standard approach here (we refer to
\cite{Henneaux:1992ig,Gomis:1995he,PS} for reviews).

\medskip

The first step is the introduction of an appropriate number of ghost
fields. More precisely, independent ghost fields are associated to
each element of the generating set of gauge transformations of the
theory, ghosts-for-ghosts for each element of the generating set of
reducibility operators, additional ghosts for the reducibility
operators of the reducibility operators, and so on. The collection of
original gauge and matter fields and of all the ghosts will be called
fields and denoted $\f^A$ in the following.

\medskip

For each field $\phi^A$, one introduces an antifield $\phi^*_A$, which
allows for a definition of an odd graded Lie bracket, the antibracket, in the
space of functionals of fields and antifields\footnote{We use DeWitt's
condensed notation.}, \bea (\cdot,\cdot)=\vddr{\cdot}{\phi^A}
\vddl{\cdot}{\phi^*_A}-\vddr{\cdot}{\phi^*_A}.
\vddl{\cdot}{\phi^A}\eea where right  and left  derivatives are defined
by $\delta F = (\delta^R F/\delta z^\a) \delta z^a = \delta z^a
(\delta^L F/\delta z^\a)$.
\medskip

Using the antifields, one couples to the original gauge invariant
action the gauge transformations with gauge parameters replaced by
ghosts, and also the various reducibility operators with their
parameters replaced by ghosts-for-ghosts. This action can then be
completed to the so-called solution of the master equation
$S[\phi,\phi^*]$ at ghost number $0$ satisfying\bea 
(S,S)=0.\label{me} \eea The associated BRST differential $s$ is
canonically generated in the antibracket by the solution of the master
equation, $s=(S,\cdot)$; it  acts in particular on both the fields and
the antifields.

\medskip

For gauge-fixing purposes, a non-minimal sector is added to $S$.
Usually, it consists of adding auxiliary fields coupled to the
antifields of the antighosts. Then one chooses a gauge-fixing fermion
$\Psi$ that depends only on the fields of the original and the
non-minimal sector. After performing a
canonical transformation generated by $\Psi$, 
the solution of the master equation has an invertible kinetic
part for the fields and can be used as a starting point for
perturbative quantization. Formal path integral arguments then show
the independence of correlation functions of BRST invariant operators
on the choice of the gauge-fixing fermion. One of the main problems is
to investigate to what extent the same property holds true in the
renormalized theory.

\medskip

In the power-counting renormalizable case, general results imply that
the quantum counterpart of (\ref{me}) can be expressed in terms of the
generating functional $W[j,\phi^*]$ of connected Green functions in
the presence of antifields\footnote{In the case of Yang--Mills gauge
theories, the antifields that appear in the gauge fixed action
correspond to sources for the BRST variations of the fields. From the
point of view of renormalization, they are needed because not only the
terms in the action, but also the non-linear BRST transformations, are
subject to quantum corrections.} as the anomalously broken
Slavnov--Taylor identity \bea j_A\vddl{W}{\phi^*_A}=\hbar \langle {\cal
A} \rangle,\label{asti}\eea with ${\cal A}$ a local integrated functional of
ghost number $1$. Equivalently, when expressed in terms of the
generating functional $\G[\phi,\phi^*]$ of 1PI Green functions, the
anomalous Slavnov--Taylor identity (\ref{asti}) becomes the anomalous
Zinn-Justin equation \bea \half (\G,\G)=\hbar{\cal A}\circ
\G. \label{azje} \eea The appropriate form of gauge invariance that a
renormalizable theory should satisfy at the quantum level is the
non-anomalous Slavnov-Taylor identity, or equivalently the
non-anomalous Zinn-Justin equation.

\medskip

The anomalous breaking ${\cal A}\circ \G$ satisfies the consistency
condition \bea (\G,{\cal A}\circ \G) =0\label{acc}, \eea which implies
to lowest non-trivial order in $\hbar$ that the anomaly ${\cal A}$
satisfies the consistency condition $s{\cal A}=0$. Because BRST exact
anomalies can be absorbed by BRST breaking counterterms at ghost
number $0$, a sufficient condition for the absence of the lowest order
anomaly is the vanishing of the BRST cohomology $H^1(s)$ in the space
of local functionals of fields and antifields at ghost number $1$.

\medskip

Suppose now that there are no non trivial anomalies and that, to
lowest order, the trivial anomalies have been canceled by BRST
breaking counterterms.  If one wants to preserve the non-anomalous
Zinn-Justin equation to that order, the remaining finite counterterms
$C$ must be BRST invariant, $sC=0$.
(In a regularization-renormalization approach, it follows from
(\ref{azje}) that the lowest order divergences $D$, which are
integrated local functionals in ghost number $0$, are also required to
be BRST invariant, $sD=0$). BRST exact counterterms can be absorbed by
canonical field-antifield redefinitions. It can then be shown that a
sufficient condition that allows for an iterative proof of the
renormalizability of the theory, while preserving the non-anomalous
Zinn-Justin equation, is the property of stability, namely that, to each
element of the BRST cohomology $H^0(s)$ in the space of local
functionals of fields and antifields at ghost number $0$, there
corresponds an independent coupling constant in the action.

\medskip

Note that from the point of view of Green functions, the antifields
are merely an (extremely efficient) bookkeeping device, which  tell
how certain combinations of Green functions with certain insertions of
operators should be renormalized.

\section{Cohomological Considerations}

In this section, we discuss local BRST cohomology with antifields
included (see \cite{Barnich:2000zw} for a review and references).

In the renormalization of gauge theories, an important part  is played
by local BRST cohomology or, more precisely, by the cohomology groups
of the BRST differential $s$ at ghost number $0$ and $1$ in the space
$\cF$ of local functionals of the fields and the antifields. Even
though the BRST differential that arises directly in the problem of
renormalization is the one associated to the non-minimal and
gauge fixed solution of the master equation $S$, the crucial point is that
the cohomology is isomorphic to the canonical, gauge invariant BRST
cohomology of the minimal, non-gauge fixed solution of the master
equation.

The information contained in the BRST cohomology groups involving the
antifields can be described entirely without these antifields.
Indeed, consider an expansion of the BRST differential in canonical
form according to the canonical antifield number, which consists of
assigning degree $0$ to the fields, degree $1$ to the antifields of
the original gauge and matter fields, degree $2$ to the antifields of
the ghosts, degree $3$ for those of the ghosts for ghosts, etc., so
that $s=\delta+\gamma+s_1+\dots$. Here $\delta$ lowers the antifield
number by $1$ and is related to the original gauge invariant equations
of motion.  The operator $\gamma$ is of degree $0$ and its action on
the gauge and matter fields, after putting the antifields to zero,
corresponds to gauge transformations with gauge parameters replaced by
ghosts, while its action on the fields of the non-minimal sector is
trivial.  Finally, the operators $s_k$ are of degree $k$. It turns out
that at all positive ghost numbers $g$, the BRST cohomology in $\cF$
is isomorphic to the cohomology of the differential $\gamma$ in the
space of local on-shell functionals $\cF_W$ depending on the fields
alone: \bea H^g(s,\cF)\simeq H^g(\gamma,\cF_W),\ g\geq 0.  \eea By
on-shell functionals, we mean equivalence classes of functionals
modulo functionals that vanish when the gauge invariant equations of
motion of the original theory hold.

Furthermore, at ghost number $-1$, the BRST cohomology in $\cF$ is
isomorphic to the space of equivalence classes of global symmetries of
the original gauge invariant theory, where two global symmetries are
equivalent if they differ on-shell by a gauge symmetry. This last
space turns out to be isomorphic to the space of equivalence classes
of conserved currents modulo divergences of superpotentials and
on-shell vanishing currents.  For irreducible gauge theories, at ghost
number $-2$, the BRST cohomology in $\cF$ is isomorphic to the space
of equivalence classes of global reducibility identities, i.e. to
classes of field dependent parameters of gauge transformations that
make the corresponding gauge symmetries vanish on-shell, modulo
on-shell vanishing gauge parameters. This space is, in turn, isomorphic
to the space of equivalence classes of conserved superpotentials
modulo divergences of skew-symmetric rank $3$ contravariant tensors
(depending locally on the original gauge and matter fields and their
derivatives) and on-shell vanishing superpotentials.  The BRST
cohomology (for irreducible gauge theories) in $\cF$ in ghost number
less than $-2$ vanishes.

Because it is the gauge fixed form of the BRST symmetry that arises
naturally in the problem of renormalization, it is sometimes useful to
describe the content of the local BRST cohomology groups in terms of
the gauge fixed theory, without the associated antifields. In the
gauge fixed form, the appropriate expansion of the differential $s$ is
according to the gauge fixed antifield number, which consists of
assigning antifield number $0$ to the fields and antifield number $1$
to all the antifields, so that
$\tilde{s}=\delta^g+\gamma^g+\l^g+\dots$. In this expansion $\delta^g$
lowers the gauge fixed antifield number by $1$ and is related to the
gauge fixed equations of motion including the fields of the
non-minimal sector. The operator $\gamma^g$ depends on the choice of
gauge fixing and is of gauge fixed antifield number $0$. Its action on
the fields, after putting to zero the antifields, is often referred to
as the BRST symmetry, because it corresponds to a rigid, global
symmetry of the gauge fixed action.  For generic gauge theories, this
symmetry is only nilpotent on-the-gauge-fixed-shell. Finally, the
operator $\l^g$ is of gauge fixed antifield number $1$, and the dots
indicate operators of higher gauge fixed antifield number.

\medskip

The cohomology classes corresponding to the operator $\gamma^g$
naturally arise in an operator formalism like the one we discuss in
section \ref{qnm}. We discuss the connection between the gauge fixed and
gauge invariant cohomology in detail in the next section.

\section{Gauge Invariant and Gauge Fixed Cohomology}

Consider a solution $S[\f, \f^*]$ of the master equation. The gauge
invariant cohomology is given by solutions of
$(S[\phi,\phi^*],A[\phi,\phi^*])=0$ up to solutions of the form
$A=(S,B)$, where $A$ and $B$ are local functionals of the fields and
antifields. The equivalence classes $[A]$ are independent of the
variables of the non-minimal sector $B,\bar C$ and their antifields,
and there is no reference to any gauge fixing.  Furthermore, the gauge
invariant classes $[A]$ can be described in terms of antifield-independent
quantities: they are completely determined by classes
$[A_0]$ satisfying $\gamma A_0\approx 0$, where trivial solutions are
given by $A_0\approx \gamma B_0$.

The consistency condition that arises in standard
approaches to renormalization is $(\tilde S[\phi,\tilde\phi^*],\tilde
A[\phi,\tilde\phi^*]\4)=0$ with $\4A$ of ghost number $1$ for the
anomalies and of ghost number $0$ for the counterterms, where
$\4S[\phi,\4\phi^*]=S(\phi,\4\phi^*+\vdd{\Psi}{\phi})$ is the
gauge fixed action where  antifields are present.
Here and in what follows tilded quantities  refer to the gauge fixed theory.
Solutions
of the form $\tilde A=(\4S,\4B\4)$ are trivial in the following sense:
for $g=1$, the anomaly $\4A$ can be absorbed by adding to the
Lagrangian the BRST breaking counterterm $-\4B$ and, for $g=0$, the
counterterm $\4 A$ can be absorbed by a canonical field/antifield
redefinition.  Because the presence of the antifields allows an
implementation of  the gauge-fixing as a (canonical) change of variables
and because the non-minimal sector is cohomologically trivial, it can be
shown that the equivalence classes of solutions
$[\4A]$ are isomorphic to solutions $[A]$ of the gauge invariant
cohomology.

The gauge fixed cohomology groups for any ghost number have been worked out in
appendix A of \cite{Barnich:1999cy} (see also \cite{Henn2}). It is shown 
that  the BRST cohomology in $\cF$ at ghost number $g$ is isomorphic to the
direct sum of two spaces:
\begin{itemize}
\item The first space is a linear subspace of  $H^g(\gamma^g,\cF_W^g)$,
where $\cF^g_W$ is the space of on-shell functionals (for the gauge
fixed field equations). Each element of $H^g(\gamma^g,\cF_W^g)$
determines a first order deformation
of the gauge fixed action and the gauge fixed BRST symmetry
and the subspace is determined by requiring that
the deformed BRST symmetry is nilpotent (on the
gauge fixed shell). An example of an element  of $H^g(\g^g,\cF_W^g)$
that does not fulfill this condition, and thus does not correspond to
a local BRST cohomology class in $\cF_W^g$ is the Curci--Ferrari mass term
\cite{Ferrari,CuFe}.
\item The second space completing $H^g(s,\cF)$ 
is isomorphic to the space of local BRST cohomology
classes in $\cF$, that, when expressed in terms of the gauge fixed
variables and after putting the antifields to zero, become $\gamma^g$
exact on the gauge fixed shell. This space can also be described in
terms of conserved currents at ghost number $g{+}1$ as we discuss
below.
\end{itemize}

Our starting point is
(A.3) of \cite{Barnich:1999cy}, which we explicitly spell out
below. Recall that $\tilde{s} =\d^g + \g^g + \l^g + ...$ is the gauge
fixed BRST differential. The nilpotency of $\tilde{s}$ implies \be
(\d^g)^2=0, \qquad \d^g \g^g + \g^g \d^g =0, \qquad \d^g \l^g + \l^g
\d^g + (\g^g)^2=0 \label{nil}. \ee The cocycle condition for the first
subspace is given by \bea
\gamma^g \4A_0+\delta^g \4A_1=0, \label{11} \\ \lambda^g
\4A_0+\gamma^g \4A_1 +\delta^g \4A_2=0,
\label{1} \eea
while the coboundary condition is \be \tilde{A}_0 =\g^g \tilde{B}_0 +
\d^g \tilde{B}_1.\label{cob1} \ee Quantities that are denoted by a
capital letter always refer to integrated quantities. Their subscripts
here and in what follows denote the (gauge fixed) antifield number.
Eqs. (\ref{11}) and (\ref{cob1}) show that $[\tilde{A}_0]$ is an
element of $H(\g^g,\cF_W^g)$. It was shown in detail in
\cite{Barnich:1999cy} that the condition (\ref{1}) is equivalent to
the nilpotency of a deformed BRST charge.

Let $j_{0}^\mu$ be the BRST  current that generates the
transformations $\g^g \f^A$. Let also $\tilde{A}_0 = \int d^n x\,
\cl_1$ (the reason for the notation will be become clear in a
moment) and $\tilde A_1=-\int d^nx\,\Delta\phi^A\tilde\phi^*_A$.
The un-integrated on-shell version of (\ref{11}) and (\ref{1}) is
\bea \gamma^g\phi^A \vddl {\cl_1}{\phi^A}\approx^g \6_\mu
j^{\mu}_1,
\label{77} \\
J^{(0)\mu A} (\vddl{\cl_1}{\phi^A})+\D j_{0}^\mu + \gamma^g
j_{1}^{\mu}\approx^g \6_\nu T^{[\nu\mu]} \label{7},\eea  
where $J^{(0)\mu A}(\cdot)$ may involve spacetime derivatives 
acting on the quantity inside the parenthesis. 
The coboundary condition becomes \be \label{cob2}
\cl_1 \approx^g \g^g b_0 + \pa_\mu k_0^\mu\,  . \ee Eq. (\ref{77})
follows straightforwardly from (\ref{11}) by noting that $\d^g$
acting on antifields yields the (gauge fixed) field equations. Eq.
(\ref{77}) implies that $(j_0^{\mu} + e j_1^{\mu})$ is a conserved
current of the theory with Lagrangian $\cl_0 + e \cl_1$, where
$\tilde{S}(\f, \f^*{=}0)=\int d^n x\, \cl_0$. Let $(\g^g + e \D)
\f^A$ be the transformation rules generated by the modified BRST
current. Relation (\ref{7}), or equivalently (\ref{1}), expresses
the fact that the modified BRST charge is nilpotent, $ (\gamma^g +
e \Delta)^2 \approx' O(e^2)$ (see 
\cite{Barnich:1999cy} for the details). Eq. (\ref{cob2}) says that
trivial solutions $\cl_1$ are the ones that are weakly BRST exact,
up to  a total divergence. That $\cl_1$ of the form of
(\ref{cob2}) indeed satisfies (\ref{77}) and (\ref{7}) is most
easily seen on the equivalent formulations (\ref{11}) and
(\ref{1}). Furthermore, by a (somewhat involved) computation one 
can show that (\ref{7}) is equivalent to
\bea \vec{k}^{A\mu } (\vddl{\cl_1}{\phi^A})+2 \gamma^g j^\mu_1
\approx^g \6_\nu T^{\prime[\nu\mu]} \label{78},\eea where
$\vec{k}^{A\mu}$ may contain derivatives with respect to the
fields and their derivatives, i.e. terms of the form
$k^{AB\mu}\frac{\partial}{\partial \phi^B}(\vddl{\cl_1}{\phi^A})$
etc.

The cocycle condition for the second subspace reads \bea \delta^g
\4C_1=0, \label{22} \\ \gamma^g \4C_1+\delta^g \4C_2=0\, .
\label{2}\eea In un-integrated form, these read \bea \6_\mu \5
l^{\mu}_0\approx^g 0, \label{cur} \\ \gamma^g \5
l^{\mu}_0\approx^g \6_\nu S^{[\nu\mu]} .\label{8} \eea Eq.
(\ref{cur}) is a straightforward rewriting of (\ref{22}), while
(\ref{8}) is derived by a standard application of descent equation
techniques: we apply $\d^g$ on (\ref{2}) and use (\ref{nil}) and
(\ref{22}) and the fact that $d$ is acyclic. Notice that these
conditions imply that $\bar{l}^\m_0$ is a  BRST invariant
conserved current. Furthermore, given a particular solution
$(\cl_1,j^\mu_1)$ of equations (\ref{77}) and (\ref{78}),  the
remaining ambiguity in $j^\mu_1$ is precisely the general
solution $\bar l^\mu_0$ of (\ref{cur}) and (\ref{8}), i.e.,
$(\cl_1,j^\mu_1+\bar l^\mu_0)$ satisfies (\ref{77}) and (\ref{78})
iff $\bar l^\mu_0$ solves (\ref{cur}) and (\ref{8}).

The coboundary condition is \bea
0=\gamma^g \4D_0+\delta^g \4 D_1\, ,  \label{55} \\ \4 C_1=\lambda^g \4D_0
+\gamma^g \4D_1+\delta^g\4D_2\, ,
\label{5}\eea which, in un-integrated form, reads
\bea
\6_\mu \rho^\mu_0 \approx^g -\gamma^g \4 d_0, \label{10'} \\ \5
l^{\mu}_0\approx^g -\gamma^g \rho^\mu_0+\6_\nu R^{[\nu\mu]}\, .
\label{10}\eea The derivation of these equations is similar to the
ones in (\ref{cur}) and (\ref{8}) (to derive (\ref{10}) act on (\ref{55})
by $\g^g$ and on (\ref{5}) by $\d^g$  and use (\ref{nil})).

In summary, the cocycle condition of the gauge fixed cohomology,
\be (\tilde S,\tilde A\4)=0, \qquad \tilde A=\4A_0+(\4A_1 +\4C_1)
+(\4A_2+\4C_2)+\dots \ee is equivalent to
(\ref{11}),(\ref{1}),(\ref{22}),(\ref{2}), or in un-integrated
form to (\ref{77}), (\ref{7}), (\ref{cur}), (\ref{8}), and the
coboundary condition \be \4A=(\4S,\4B\4), \qquad
\4B=(\4B_0+\4D_0)+(\4B_1+\4 D_1)+\dots \ee is equivalent to
(\ref{cob1}),(\ref{55}),(\ref{5}), or in un-integrated form to
(\ref{cob2}), (\ref{10'}), (\ref{10}). We thus see that $[\tilde
A]$ is equivalent to $([\4 A_0],[\4C_1])$, respectively to
$([\cl_1],[\bar l^\mu_0])$. This should be contrasted with the
case of the gauge invariant cohomology, where the antifield
independent part $[A_0]$ was sufficient to completely determine
$[A]$.

\section{An Example}

In this section we illustrate with a simple example
the discussion in the previous section. In particular
we demonstrate the emergence of non-trivial currents in specific gauges.
This example was also briefly discussed in \cite{Barnich:1999cy}.
We consider Maxwell theory with a free massive scalar
$\F$. The gauge invariant action is
\be
S_0 = \int d^4 x \left(-{1 \over 4} F^{\mu \n} F_{\m \n}
- \half \pa_\m \F \pa^\m \F - \half m^2 \F^2 \right).
\ee
The gauge invariant non-minimal solution of the master equation
is
\be \label{snm}
S(\f,\f^*) = S_0 + \int d^4 x\, (A^{*\mu} \pa_\mu C + \bar{C}^* B),
\ee
where $C$ is the ghost field, $\bar{C}$ is the antighost field
and $B$ is the BRST auxiliary field.
In the gauge invariant theory, $\int d^4 x\, \F$ is BRST closed,
$(S, \int d^4 x\, \F)=0$, but not BRST exact, so $[\int d^4 x\, \F]$
is a non-trivial class of $H^{0,n}(s|d)$.

We now consider fixing the gauge using the gauge fixing condition
$\pa^\m A_\m = \mu \F$, where $\m$ is a mass parameter.
The limit $\mu \to 0$ yields the well-known Lorentz gauge.
The gauge fixing fermion that implements this gauge is given by
\be
\Psi = \int d^4 x\, \bar{C} (\half \a B + \pa_\mu A^\mu + \mu \F).
\ee
The gauge fixed action $S^g = S(\f, \tilde{\f}^* + \d \Psi/\d \f)$ is
equal to
\bea \label{gf}
S^g &=& \int d^4 x \left( -{1 \over 4} F^{\mu \n} F_{\m \n}
- \half \pa_\m \F \pa^\m \F - \half m^2 \F^2 - \pa^\m \bar{C} \pa_\m C
+(\half \a B + \pa_\mu A^\mu + \mu \F) B \right. \nonumber \\
&& \left. \qquad \qquad
+ \tilde{A}^{*\mu} \pa_\mu C + \tilde{\bar{C}}^* B \right).
\eea

We now set the antifields to zero.
As is well known, the BRST variation of the antighost yields the gauge fixing
condition,
\be \label{antigh}
\g^g \bar{C} \approx^g - {1 \over \a} (\pa^\m A_\m + \mu \F),
\ee
where the field equation of $B$ was used. Integrating we get
\be \label{ntrph}
\mu \int d^4x\, \F \approx^g \g^g \int d^4x\, \a \bar{C}.
\ee
The action (\ref{gf}) is invariant under constant shifts of the
antighost, $\d \bar{C}=\e$. The corresponding Noether current
is given by
\be
\bar{l}_0^\mu = \pa^\mu C = \g^g A^\m.
\ee
This current is BRST invariant,
\be
\g^g \bar{l}_0^\mu = 0.
\ee

Let us first consider the case $\mu=0$, i.e. the standard Lorentz gauge.
In this case, the current $\bar{l}_0^\mu$ is trivial since the condition
in (\ref{10})
are satisfied with $\r_0^\m=-A^\mu$ and
$\tilde{d}_0 = - \bar{C}$. Furthermore, $\int d^4x\, \F$ is non-trivial,
as in the gauge invariant formulation (the relation
(\ref{ntrph}) just says that $\int d^4 x\, \bar{C}$ is BRST closed).

We now discuss the case $\m \neq 0$. The relation (\ref{ntrph}) implies
that $ \int d^4x\, \F$ is trivial. The current $\bar{l}_0^\mu$, however, is
no longer trivial because the first eq. in
(\ref{10}) is not satisfied due
to the last term in (\ref{antigh}). We thus see that in this gauge
we lost one non-trivial class at the level of $\cl_1$ but gained another
one at the level of $j^\m_1$.

There is an elegant reformulation of this discussion
when antifields are present.
As discussed, $\int d^4 x\, \F$ represents a non-trivial class
in the gauge invariant formulation.
Furthermore, the functional
$\int d^4 x\, \bar{C}^*$ has ghost number zero, is BRST closed,
$(S, \int d^4 x\, \bar{C}^*)=0$,
but it is also BRST exact, $\int d^4 x\, \bar{C}^*=(S,\int d^4 x\, B^*)$,
and thus it represents the trivial class. Notice that
$\int d^4 x\, \bar{C}^*$ generates (by the antibracket)
arbitrary constant shifts of the antighost $\bar{C}$.
This a trivial symmetry because
it comes from a gauge symmetry: the action (\ref{snm}) is invariant
under local shifts of the antighost.

Let us now consider the gauge fixed case. A simple computation
yields
\be
\mu \int d^4x\, \F = (S^g, \int d^4 x\, (\tilde{B}^* + \a \bar{C})) -
\int d^4x\, \tilde{\bar{C}}^*.
\ee
We thus see that when $\mu=0$, $\int d^4x\, \tilde{\bar{C}}^*$ is
trivial as in the gauge invariant formulation, but when $\mu \neq 0$
both $\int d^4x\, \F$ and $\int d^4x\, \tilde{\bar{C}}^*$ represent the
same cohomology class.

We would like to emphasize that the discussion so far was about
possible counterterms in the quantum effective action (and, as we discuss
in the last section, about the local normalization ambiguity in the 
BSEG approach). There is a separate issue concerning
quantum operators and whether they are trivial inside correlation functions,
as we now discuss. To analyze this question we examine  
integrated correlation functions of $d$-exact operators, 
$\pa_\mu K^\mu$, with an arbitrary number of gauge invariant operators $O_i$,
$\int d^4 x\, \langle T [(\pa_\mu K^\mu)(x) O_1(x_1) \cdots O_n(x_n)] \rangle$.
Of course, for the question to be well-posed the correlators should be free 
of IR divergences. Let us consider for concreteness the case of 2-point 
functions. Standard manipulations lead to 
\be \label{exm}
\int d^4 x\, \langle T [(\pa_\mu K^\mu)(x) O_1(y)] \rangle
= - \int d^3 x \langle[K^0(\vec{x},t), O_1(\vec{y},t)] \rangle 
+\int d^4 x\, \pa^x_\mu \langle T [K^\mu(x) O_1(y)] \rangle
\ee
The first term is standard, but the second term is usually dropped
because it is the integral of a total derivative and it is usually
assumed that all fields fall off sufficiently fast at infinity
so that such terms can be ignored. It is such terms that are the focus 
of our analysis here. 
  
Let us now consider $K^\mu = A^\mu$ and $O_1=\Phi$. In this case 
it is straighforward to compute explicitly all quantities.
Because of the gauge-fixing there is an off-diagonal propagator,
\be
\langle T [A^\mu(x) \F(y)] \rangle = \mu \int \frac{d^4 p}{(2 \p)^4} 
e^{i p \cdot (x-y)}  {p^\mu \over p^2 (p^2 + m^2)}\,.
\ee
It follows that 
\be \label{res}
\int d^4 x\, \pa^x_\mu \langle T [A^\mu(x) O_1(y)] \rangle 
= i {\mu \over m^2}\,.
\ee
We thus find that in this specific example a correlator that is naively 
zero because it is the integral of a total derivative is actually 
non-zero. 

Let us try to understand this result.
Standard arguments imply that when an operator is BRST exact
its correlation functions with any gauge invariant operators $O_i$
vanishes,
\be \label{trivial}
\langle T[(\g^g A)(x) O_1(x_1) \cdots O_n(x_n)] \rangle =0,
\ee
where $A$ is any local operator.
We can use this Ward identity with $A=\bar{C}$ to show
(the well-known fact) that all correlation
functions of the gauge fixing relation with gauge invariant
operators vanish. In our case this implies,
\be \label{cor}
\mu \langle T[\F(x) O_1(x_1) \cdots O_n(x_n)] \rangle
=- \langle T[(\pa_\m A^\m)(x) O_1(x_1) \cdots O_n(x_n)] \rangle.
\ee
This relation tells us that the correlation functions of the
{\it gauge variant} operator $\pa_\mu A^\mu$ are determined
in this gauge by the correlation functions of the
{\it gauge invariant} operator $\F$. Notice that
the correlation functions $\langle T[\F(x) O_1(x_1) \cdots O_n(x_n)] \rangle$
are the same in all gauges (since they involve gauge invariant
operators) but the correlation functions of
$\langle T[\pa_\m A^\m(x) O_1(x_1) \cdots O_n(x_n)] \rangle$
depend on the gauge under consideration. 

We now consider an integrated version of (\ref{cor}) with $O_1=\F$ 
and no other operators,
\be \label{corgf}
\mu \int d^4 x\, \langle T[\F(x) \F(y)] \rangle
=- \int d^4 x\, \langle T[(\pa_\m A^\m)(x) \F(y)] \rangle.
\ee
First, the equal-time commutator between $A^0$ and $\F$ is zero
(as an explicit computation shows)
so one can freely move the derivative outside the time-ordering in the 
right hand side. The l.h.s. is equal to 
\be
\mu  \int d^4 x\, \langle T[\F(x) \F(y)] \rangle 
= -i \mu 
\int d^4 x {d^4 p \over (2 \pi)^4} {e^{i p \cdot (x-y)} \over p^2 +m^2}
= - i {\mu \over m^2}
\ee
so (\ref{corgf}) yields (\ref{res}).

As mentioned above, ordinarily integrals of total derivatives of 
correlators are set to zero because all fields are assumed to fall 
off sufficiently fast at infinity. The asymptotic behavior of 
$A_\mu$, however, depends on the gauge. 
In the current case the gauge fixing condition implies, 
\be
\int_{\pa M} d \s^n\, A_n = - \mu \int_M d^4 x\, \F,
\ee
which is non-zero. This (classical) statement is implemented in the 
quantum theory via the Ward identity (\ref{cor}). To 
take into account such non-trivial boundary conditions one may 
consider the cohomology of $d$ with compact support, i.e. 
a $d$-closed element is considered trivial if it is $d$-exact 
and vanishes at infinity. The operator $\pa_\mu A^\mu$ in thus
non-trivial in $H_{compact}(d)$.
 
The existence of currents of the form of the second eq. in (\ref{10})
that are non-trivial because the first condition in
(\ref{10}) is not satisfied always implies that there are such
non-trivial $d$-closed operators. Indeed, the conservation
of $\bar{l}_0^\mu$ in the second condition in (\ref{10}) implies only that
\be \label{nontr}
\g^g r \approx^g 0,
\ee
where $\pa_\mu \r_0^\mu \approx^g r$. Now consider the correlation
function of $\pa_\mu \r_0^\m$ with arbitrary number
of gauge invariant operators. Naively, this should be equal to zero
but because of (\ref{nontr}) we obtain
\be
\int d^4 x\, \langle T[(\pa_\m \r_0^\m)(x) O_1(x_1) \cdots O_n(x_n)] \rangle =
 \int d^4 x\, \langle T[ r(x) O_1(x_1) \cdots O_n(x_n)] \rangle,
\ee
which is (generically) non-zero since the right hand side
is a correlation function of gauge invariant operators. In contrast, when the
current is trivial, $r$ is $\g^g$-exact and the correlation
function on the right hand side vanishes.

These considerations indicate that at the level of operators
there is an alternative
cohomology analysis to the one presented here and in the
appendix of \cite{Barnich:1999cy} where the non-trivial
currents are due to $d$ having a non-trivial cohomology.

\section{Quantum Noether Method} \label{qnm}
\setcounter{equation}{0}

As already mentioned in the introduction, 
a construction of theories with global (=rigid)
and/or local symmetries within the
Bogoliubov-Shirkov-Epstein-Glaser (BSEG) operator approach was
presented in \cite{paper1,proc}. 
Recall that in this approach the $S$-matrix is constructed
perturbatively in the asymptotic Fock space as a formal
power series
\begin{equation}
\label{GC}
 S(g) = 1 + \sum_{n=1}^\infty \frac{1}{n !}  \int dx_1^4 \cdots dx_n^4
\,  T_n(x_1, \cdots, x_n; \hbar) \, g(x_1) \cdots g(x_n).
\end{equation}
The coupling constant $g$ is replaced by a tempered test function
$g(x) \in {\cal S}$ (i.e. a smooth function rapidly decreasing at
infinity), which switches on the interaction.  The $n$-point
operator-valued distributions $T_n \in {\cal S'}$ are the central
objects, where ${\cal S'}$ denotes the space of functionals on ${\cal
S}$.  They should be viewed as mathematically well-defined
(renormalized) time-ordered products,
\begin{equation}
 T_n(x_1, \cdots, x_n; \hbar) = T \left[T_1(x_1) \cdots
 T_1(x_n)\right],
\end{equation}
of a given specific coupling $T_1$, which is regarded as part 
of the definition of the theory (for example $T_1 = i/\hbar :\phi^4:$).  We
note that the expansion in (\ref{GC}) is {\it not} a loop expansion.
Each $T_n$ in (\ref{GC}) can receive tree-graph and
loop-contributions. One  can distinguish the various contributions
from the power of $\hbar$ that multiplies them.

We are interested in constructing theories where the $S$-matrix is
invariant under a certain symmetry operation, generated by a
well-defined operator $Q$ in the asymptotic Fock space: \be [Q,
S]=0. \label{Sinv} \ee The operator $Q$ acting on asymptotic fields
generates their asymptotic transformation rules \be [Q, \f^A \} = -i
\hbar \gamma_0 \f^A, \label{ch1} \ee where $[A, B\}$ denotes a graded
commutator. The latter are necessarily linear in the asymptotic
fields.

\medskip

We would like to carry out the construction before the so-called adiabatic
limit.  Thus, instead of working with (\ref{Sinv}), we require that
\be [Q, T_n(x_1, \ldots, x_n; \hbar) \} = \sum_{l=1}^n {\pa \over \pa
x^\m_l} T^\m_{n/l}(x_1, \ldots, x_n; \hbar)
\label{qn}
\ee holds in the distributional sense for $n \geq 1$ and for some $T^\m_{n/l}$;
 we use the abbreviation $\pa/ \pa x^\m_l =
\pa^l_\m$. For $n=1$,  Eq. (\ref{qn}) reads  \be [Q, T_1 \} =
\pa_\m T^\m_{1/1},\label{q1} \ee and imposes restrictions on the starting
point of the BSEG procedure, namely on the coupling $T_1$.\footnote{
The distributions $T^\mu_{n/l}$ in (\ref{qn}) are then given by the
$T$-products of $(n-1)$ vertices $T_1$ and one vertex $T_{1/1}$ at the
$l$th position: $T[T_1 (x_1) \cdots T_{1/1}^\mu(x_l) \cdots
T_1(x_n)]$.}  Once the coupling $T_1$ has been determined, the rest of
Eqs. (\ref{qn}) impose relations between  the constants left
unspecified by the requirement of causality and Poincar\'{e}
invariance.  This is analogous to the situation in the conventional
Lagrangian approach, where symmetry considerations restrict the
possible terms in the Lagrangian and then the corresponding symmetries
at the quantum level impose certain relations between the $Z$ factors.

\medskip

Since different non-linear transformations may have the same linear
limit, it is not {\it a priori} obvious whether a theory constructed by
BSEG satisfying (\ref{Sinv}) has any underlying non-linear structure
at all. To address this issue one can work out the precise
consequences of the operator equation (\ref{qn}) in specific models
and try to reproduce the Ward identities derived in the Lagrangian
approach, using the full non-linear transformation.~\footnote{ This
approach was followed in \cite{DHS95,H95} for the case of $SU(n)$
gauge theory in the Feynman gauge coupled to fermions where it was
shown that (\ref{qn}) implies the Slavnov-Taylor identities for
connected Green functions.} An alternative and complementary approach
is to try to find a direct correspondence between the Lagrangian
approach and the BSEG formalism.  The obvious advantage of such an
approach is that it will provide a model-independent connection
between the two approaches.

In order to establish such a connection to the Lagrangian approach, 
the symmetry constraint (\ref{qn}) was reformulated into an
equivalent, but more transparent formulation based on the Noether
method in \cite{paper1}. 
Recall that in the Lagrangian approach Ward identities are
derived using the conservation of symmetry currents inside correlation
functions. The proposal in \cite{paper1,proc} is to adopt this
approach in the operator formalism: we constrain the local ambiguity
by requiring that the corresponding Noether current be  conserved
inside correlation functions.

We start by including in the theory the coupling $g_\m j^\m_0$, where
$j^\m_0$ is the Noether current that generates the asymptotic (linear)
symmetry transformations of the fundamental fields.  Given such a
current, one can obtain a corresponding interacting field operator
$j_{0,int}^\mu$.\footnote{ The perturbation series for the interacting
field operator $j^\mu_{0,int}$ of a free field operator $j_0^\mu$ is
given by the advanced distributions of the corresponding expansion of
the $S$-matrix: \be
\label{advanced}
j_{0,int}^\mu (g,x) = j_0^\mu (x) + \sum_{n=1}^\infty \frac{1}{n!}
\int d^4 x_1 \ldots d^4 x_n Ad_{n+1} \left[T_1 (x_1) \ldots T_1 (x_n);
j_0^\mu (x) \right] g(x_1) \ldots g(x_n), \ee where $Ad_{n+1}$ denotes
the advanced operator-valued distribution with $n$ vertices $ T_1 $
and one vertex $ j_0^\mu (x) $ at the $(n+1)$th position; $g(x_k)$ is
a tempered test function, which, in the adiabatic limit, becomes the
coupling constant of the theory.}  We then propose, as a general Ward
identity \cite{proc}: 
\be \label{cond3} 
\pa_\m j^\m_{0,int} = \pa_\mu g \, \tilde{j}^\mu_{1, int} 
\ee 
where $\tilde{j}^\mu_{1, int}$ is another interacting current
whose explicit form can be found in \cite{proc} but it will
not be needed here. One of the main results of \cite{proc} is
that the interacting current at tree level is exactly equal to the
Noether current of the Lagrangian (non-linear) theory.  It follows
that in the naive adiabatic limit, $\pa_\mu g =0$, (\ref{cond3})
reduces to the standard Ward identities.  The condition (\ref{cond3})
is a formal Laurent series in $\hbar$ so this condition is actually a
set of conditions.

Using the fact that the right-hand side of (\ref{cond3}) is linear in
derivatives, one may provide an alternative Ward identity, which does
not contain any explicit reference to $g(x)$. The reformulation is
based on the distributional identity $(\sum_{i=1}^n \pa_i) \d(x_1-x_2)
\d(x_2-x_3)...\d(x_{n-1}-x_n)=0$. It follows that if one considers a
symmetrized insertion of $j_0^\mu$ (instead of (\ref{advanced})) then
the term on the right-hand of (the symmetrized version of) (\ref{cond3})
becomes proportional to a total derivative, which can then be removed by
modifying the local ambiguity in correlation functions with one
current insertion. This version of the Ward identity was presented in
\cite{paper1}. By construction, the two Ward identities imply exactly
the same conditions on all correlators with no current insertions, but
 those that have current insertions differ by the local terms just
discussed. The Ward identity (\ref{cond3}) has the advantage that
 it involves the current $j_{0,int}^\mu$ that renormalizes to become
exactly equal to the Noether current of the Lagrangian formulation (in
the symmetrized Ward identity the renormalization of $j_0^\mu$
involves the same terms,  but with different combinatorial
coefficients). Furthermore, it is more straightforward to analyse the
anomalies of (\ref{cond3}) than that of the symmetrized condition.

It was shown in \cite{paper1,proc} that
any theory with global/local symmetry that can be viewed as
deformation of a free theory
can be constructed  by using the
symmetry condition (\ref{cond3}) (or the corresponding symmetrized
condition) and the free Noether current $j_0^\m$ as a starting point.
This class of theories includes all perturbative
QFTs.  In addition, the equivalence of any theory consistently
constructed in the BSEG formalism with a Lagrangian theory was
established.  The following was shown there:
\begin{enumerate}
\item The sum of $T_1$ and the tree-level normalizations that arise
from the requirement (\ref{cond3}) coincides with the Lagrangian that
is invariant under the non-linear transformations generated
by $j_{0,int}^\m$.  This shows that
 the full non-linear structure is present in the theory.
\item The interacting current $j_{0,int}^\m$ is renormalized by
condition (\ref{cond3}) in such a way that it is, at tree level, exactly
equal to the Lagrangian Noether current.
\item The loop normalization ambiguity is fixed in the same way as at
 tree level, provided the anomaly consistency condition has only
trivial solutions. This means that the theory is stable under quantum
correction.
\end{enumerate}
It was also shown that condition (\ref{cond3}) is equivalent to
condition (\ref{qn}).  The latter guarantees the invariance of the
$S$-matrix under the corresponding asymptotic symmetry.

\medskip

The consequence of all this is that the only information one needs in
order to construct a perturbative quantum field theory with a given
global symmetry is a set of free fields linearly realizing this
symmetry.  Even the first term in the $S$-matrix, which is usually
regarded as an input in the BSEG formalism, is now derived using the
quantum Noether condition.

\medskip

Because the BSEG construction leads to the most general quantum field
theory that is compatible with causality and Poincar\'{e} symmetry, the
quantum Noether condition allows a determination of  all consistent quantum
field theories with non-linear symmetries that are compatible with the
asymptotic symmetry represented by $j_0^\m$.

\medskip

Further consistency requirements on the theory follow by considering
multicurrent correlation functions \cite{Barnich:1999cy}.  In
particular, the two-current equation is \be \label{QNC2} \pa_\m^x
T[j_0^\m(x) j_0^\n(y) T_1 (x_1) \cdots T_1(x_n)] = \pa_\m g \,
\tilde{j}^{\m \n}, \ee for some $\tilde{j}^{\m \n}$.  In the case of a
BRST symmetry, this additional constraint implements the nilpotency of
the BRST transformation in the quantum theory. More generally, 
the two-current equation implements the algebra of symmetries. 

\medskip

The proofs given in \cite{paper1,proc,Barnich:1999cy} will not be
repeated here.  We discuss in the appendix, however, the so-called
{\it off-shell} procedure.  This alternative route to fix the local
ambiguities and local breaking terms within the {\it on-shell}
formalism provides structural (model-independent) information on the
latter.  It is this additional structural information that allows us to
establish a cohomological analysis of the local ambiguities and the local
anomalies within the BSEG operator approach.

\medskip

Recently a new Ward identity was proposed within the BSEG approach
\cite{MWI1,MWI2}. The so-called master Ward identity (MWI) is presented as
a universal renormalization condition that expresses the symmetries of
the underlying classical theory. The MWI is obtained by computing
the difference,
\be \label{MWI}
\pa_{x_1}^\m T[W_1(x_1),\dots,W_n(x_n)] - T[\pa^\m W_1(x_1),\dots,W_n(x_n)],
\ee
using an algebraic equivalent of the Wick expansion formula
(called N3 in \cite{MWI1}) and then demanding that the result
is preserved by renormalization ($W_1$,...,$W_n$ are polynomials
in free fields).  We would like to emphasize
that (as is well-known) the Wick
expansion formula is not the most general solution of the BSEG
construction. Actually the authors of \cite{MWI1} already
encountered problems in reconciling (\ref{MWI}) with known
symmetries of certain models under consideration. In particular,
it was found that the interacting Noether current in the case of
massless Yang--Mills is {\it not} compatible with the MWI and
the normalization condition N3, see formulae (157-160).
Furthermore, in the case of the non-abelian Higgs model,
the distributions $T_n$ {\it fulfilling} the MWI
have to be modified due to certain
non-linear couplings introduced by the BRST symmetry.
The resulting $n$-point distributions $T_n^N$ violate the MWI and N3
as stated below formula (215).  Since the MWI does
not capture the full non-linear structure of symmetries
(as these examples show), it seems likely that the theory satisfying MWI
will have more anomalies and not be stable under quantum corrections.

\section{Gauge Fixed Cohomology and the QNM}

In \cite{paper1,proc} the constraints imposed by the quantum
Noether condition were analysed to all orders. Let $\cl_n$
denote the local normalization ambiguity of $T_n$
\be
T_n[T_1(x_1) \dots T(x_n);\hbar] = T_{c,n}[T_1(x_1) \dots T(x_n);\hbar]
+ n! {i \over \hbar} \cl_n(\hbar) \d(x_1,\dots, x_n)\, ,
\ee
where $T_{c,n}$ denotes a reference splitting  solution. At tree level,
this is defined by using the Feynman propagator in tree-graphs.
The Lagrangian is then identified with the tree-level contribution to
$\cl=\cl_0 + e \cl_1 + \dots$ \cite{paper1}. Here $\cl_0$ is the Lagrangian
of the free theory, and $e$ is the coupling constant of the theory\footnote{
We use the convention of keeping the coupling constant explicit, in which
case the adiabatic limit is $g(x) \to 1$.}. It was shown
in \cite{paper1,proc} that the condition imposed by (\ref{cond3})
on the local terms $\cl_n$ coincides with the equations
 obtained  in the classical Noether method. These considerations, thus,
lead to an BSEG construction of all theories which
are associated at the classical level with a Lagrangian
invariant under some (non-linear) symmetry.

\medskip

A particular case of theories where the construction applies
is that of gauge theories, the relevant
global symmetry being the BRST symmetry. In this case one
can obtain the most general local terms compatible with gauge
invariance using the antifield formalism.
The objective of \cite{Barnich:1999cy} was to
show that the set of solutions of the quantum Noether conditions coincides
with the set of solutions given by the antifield method.

In this case, the one-current quantum Noether condition
(\ref{cond3}) implies \be \label{1st} \g^g \cl_1 = \pa_\m \cl^\m_1
-  \D \f^A {\delta^L \cl_0 \over \delta \f^A}, \ee whereas the
two-current Noether condition (\ref{QNC2}) yields \be \label{2nd}
J_0^{\m A}( {\d \cl_1 \over \d \f^A}) + \g^g j_1^\m + \D j_0^\m =
\pa_\n T_1^{[\n \m]} + J_1^{\m A} ({\delta^L \cl_0 \over \delta
\f^A}). \ee   Eq. (\ref{1st}) defines $\D \f^A$. It was shown in
\cite{proc} that these transformations are generated by $j_1^\mu$,
and that furthermore $j_1$ is related to $\cl_1$ by \be j_1^\mu =
- \cl_1^\mu + {\pa \cl_1 \over \pa (\pa_\mu \f^A)} \g^g \f^A \, .
\ee

Because these equations coincide with (\ref{77}) and (\ref{7}),
the most general solution $(\cl_1, j_1^\mu)$ are given by the
gauge fixed BRST cocycles discussed in section 4. Thus, we see
that the QNC correctly gives the cocycle conditions of the gauge
fixed cohomology.

We now turn to the issue of the coboundary terms.
First consider the terms of the form (\ref{cob2}).
Inserting them in (\ref{GC}) we see that they
are indeed trivial: total derivatives can be
ignored in the (naive) adiabatic limit, and the
asymptotic fields satisfy their field equation.
Furthermore, BRST exact terms vanish when
acting on physical states (since the latter are BRST closed).

Let us now consider the coboundary condition (\ref{10'}) and (\ref{10}),
\bea
\6_\mu \rho^\mu_0 = -d^A {\delta^L \cl_0 \over \delta \f^A}
-\gamma^g \4 d_0, \label{new10'} \\
\5 l^{\mu}_0 = -\gamma^g \rho^\mu_0
+ d^{\mu A} {\delta^L \cl_0 \over \delta \f^A} + \6_\nu R^{[\nu\mu]} \, .
\label{new10} \eea Recall that the starting point in the BSEG
procedure is a set of asymptotic fields and their commutation
relations, or equivalently their propagators. From these data one
may construct a gauge fixed Lagrangian $\cl_0$ that yields these
propagators. As we discussed, the gauge fixed Lagrangian and the
gauge fixed BRST current are associated with a gauge invariant
Lagrangian $L$ and a corresponding BRST current. Suppose now that
we change variables and choose another gauge,  such that the
resulting gauge fixed action changes only by a total derivative.
The propagators of the gauge fixed theory remain invariant under
this combined operation, but the BRST current can change. Clearly,
the theory constructed using either the original or the new BRST
current is the same. In this sense the difference between  the two
BRST currents is a trivial current. In other words, currents that
can be removed by a combined change of gauge and of variables are
trivial. The current $\bar{l}_0^\mu$ satisfying (\ref{new10'}) and
(\ref{new10}) is such a case. Indeed, consider the change of
variables and gauge \be \f^A \to \f^A + e d^A, \qquad \Psi \to
\Psi - e \tilde{d}_0 \, , \ee where $\Psi$ is the gauge fixing
fermion. The gauge fixed Lagrangian changes by \be \cl_0 \to \cl_0
+ e d^A {\d^L \cl_0 \over \d \f^A} + e \g^g \tilde{d}_0 = \cl_0 -
e \6_\mu \rho^\mu_0, \ee where we used (\ref{new10'}). The change
of variables and of gauge entails a change in the BRST current.
The details are somewhat involved but the final answer is that the 
original current
changes by a current $e \bar{l}_0^\mu$ of the form (\ref{new10'})
and (\ref{new10}). This analysis tell us that one should include
all non-trivial currents $\bar l^\mu_0$ in the Noether procedure.

\section*{Acknowledgments}
TH  would like to thank R. Nest, F. Scheck, and  E.Vogt for
the invitation and  for the organization of a stimulating
meeting between mathematicians and physicists. KS is supported
by NWO. The work of  GB is supported in part by the ``Actions de Recherche
Concert{\'e}es'' of the ``Direction de la Recherche
Scientifique-Communaut{\'e} Fran\c{c}aise de Belgique, by a ``P{\^o}le
d'Attraction Interuniversitaire'' (Belgium), by IISN-Belgium
(convention 4.4505.86), by Proyectos FONDECYT 1970151 and 7960001
(Chile) and by the European Commission RTN program HPRN-CT00131,
in which the author is associated to K.~U.~Leuven.

\appendix

\section{Off-Shell Procedure}

The purpose  of  this appendix is to review
the so-called {\it off-shell} procedure
presented in \cite{paper1}. In order to avoid any misunderstandings,
it should be stressed that the (well-defined)
{\it off-shell} formalism
is nothing else than a shortcut to fix  the local ambiguities and local
breaking terms within the {\it on-shell} BSEG formalism \footnote{
The field operators of the BSEG construction fulfill the
free field equations and in this sense the formalism is {\it on-shell}.}.
The advantage of the {\it off-shell} procedure  is that it
provides structural (model-independent) information
on the latter, which allows us to
establish a cohomological analysis of the local normalization ambiguity.

The symmetrized  quantum Noether condition\footnote{As
mentioned in section \ref{qnm}, this symmetrized condition is
equivalent to the Ward identity (\ref{cond3}),
$\pa_\m j^\m_{0,int} = \pa_\mu g \tilde{j}^\mu_{1, int}$,
in its implications on
physical correlation functions with no current insertions (see
\cite{proc} for an explicit proof). The Ward identity (\ref{cond3})
can be written in terms of $T$-products as
\be
\partial_\mu^x T_{n+1} \Big[ T_1(x_1) \ldots T_1(x_n); j_0^\mu (x) \Big]
= -\sum_{j=1}^{n} T_n \left[ T_1(x_1) \ldots
\widehat{T_1(x_j)} \ldots T_1(x_n);
\tilde{j}_1^\mu (x) \right] \partial_\mu^{x_j}
\delta (x_j - x)\, . \label{condt}
\ee
These distributions get smeared out by $ {g(x_1) \ldots g(x_n) \tilde {g}(x)}$, where the test-function $\tilde{g}$ differs from $g$.
One easily verifies that the symmetrization in {\it all}
variables (smearing out (\ref{condt}) by $ g(x_1) \ldots g(x_n) g(x)$)
just leads to the symmetrized quantum Noether condition (\ref{cond}).
As is described in the main text, the two Ward identities lead
to a different normalization of correlators with current insertions,
but one can readily obtain the normalization needed to satisfy
(\ref{cond}) from the normalization needed to satisfy (\ref{cond3})
(and vice versa) using the fact that the former is a symmetrized
version of the latter. The explicit local normalization
terms at order $n$ of the one-current correlators are given explicitly
in \cite{proc} for both Ward identities.} is given by
\cite{paper1}:
\be \label{cond}
\pa_\m \cj_n^{\m} (x_1, \cdots, x_n; \hbar)=
\sum_{l=1}^n \pa_\m^l \cj_{n/l}^{\m} = 0,
\ee
where
\be
\cj^\m_{n/l}=T[T_1 (x_1) \cdots j_0^\m(x_l) \cdots T_1(x_n)].
\ee
We use the abbreviation $\pa/ \pa x^\m_l = \pa^l_\m$.
For $n=1$ we have the condition  $\cj^\m_1(x_1)=j_0^\m(x_1)$.

We start the quantum Noether construction \cite{paper1}
by noting that having satisfied our fundamental
quantum Noether condition (\ref{cond}) for all $m<n$,
 Eq. (\ref{cond}) at the $n$th order
can be violated by a local distribution $A_n(\hbar)$
(which we shall call local breaking term) only:
\be \label{cons3}
\pa_\m \cj_n^{\m} (x_1, \cdots, x_n;\hbar)= \sum_{l=1}^n
\pa_\m^l \cj_{n/l}^{\m} = A_n(\hbar).
\ee
The conventional inductive BSEG construction may be applied
 to work out the consequences of
(\ref{cond}) and to fix the local terms $A_n(\hbar)$ in (\ref{cons3}).
 However, there is an alternative
route which was worked out in great detail in  \cite{paper1}, and this
under the name ``off-shell formulation of the inductive hypothesis''.
To understand how this {\it off-shell} formulation simplifies the
calculation of local {\it on-shell} terms $A_n(\hbar)$
arising from tree-level contractions, the traditional way to do  such a
calculation should first be described in more detail:
In order to obtain the local terms,
one first constructs~\footnote{ \label{nat}
$T_c$ denotes the natural splitting
solution, i.e. the Feynman propagator is used in tree-graphs.}
 $T_{c,n}[j^\mu_0(x_1)T_1(x_2)...T_1(x_n)]$, differentiates with respect
to the variable   of the current and symmetrizes in all variables.
$T_{c,n}[j^\mu_0(x_1)T_1(x_2)...T_1(x_n)]$
involves many terms and there will be a large number of cancellations
after differentiating and symmetrizing.
In particular, one already knows from
Eq. (\ref{cons3}) that all non-local
terms will cancel one another.
So the idea (which gets implemented with the help
of the {\it off-shell} trick) is to concentrate
on possible local terms, anticipating the
cancellation of all non-local terms.

By induction hypothesis, one has for all $m<n$,
\be
\sum_{l=1}^m \pa^l_\mu \cj_{m/l}^{\m} = 0 \,.
\label{offshell0}
\ee
However, the assumption that the quantum Noether method works successfully
means that there exist local normalizations such that
(\ref{cond}) is satisfied when the field equations are satisfied.
Therefore one rewrites --just as a trick-- the induction hypothesis
(\ref{offshell0}) by relaxing the field equations of the fields
$\phi^A$ (which will be  denoted by $ \ck_{AB} \f^B$ in the
following):
\be \label{offshell}
\sum_{l=1}^m \pa^l_\mu \cj_{m/l}^{\m} =
* R^{A;m}(\hbar) \ck_{AB} \f^B \d^{(m)} * , \quad m < n \, .
\ee
It should be stressed again that the {\it off-shell}
terms  on the right-hand side  are understood just as
a calculational device  to  fix the local {\it on-shell}
terms $A_n(\hbar)$ out of tree contractions
in the next inductive step $(n-1) \rightarrow n $. In the following,
the stars * are always used  within the
exact equations of the BSEG construction, as in (\ref{offshell}),
in order to
indicate these off-shell terms.
Note that all these terms are just zero
within the BSEG {\it on-shell} formalism.
Nevertheless, the {\it off-shell} terms have a well-defined meaning
and can be constructed straightforwardly with the help of BSEG
quantities. The uniqueness of these terms will be
discussed below.

\medskip

The crucial point of the construction is  that
exactly those terms that are
proportional to the field
equations are the only source of the local {\it on-shell}
terms $A_n(\hbar)$
in the next step of the induction within tree-level contractions:
at order $n$ the so-called {\it relevant contractions}, namely the
contractions between the $\f^B$ in the right-hand side of (\ref{offshell})
and $\f$ within another local term, lead to the local terms.
This implies, in particular, that no local term arises from
terms in the causal distribution at the order $n$  that
are products of more than two $T$ products.
This is in accordance with the diagrammatic picture of creation
of local terms that was discussed in \cite{paper1}.
In this manner one gets the following general formula for the
local {\it on-shell} term $A_{c,n}$  arising through tree-level
contractions at level
$n$ \cite{paper1}:
\be \label{loc}
A_{c,n}(\mbox{tree}) =  \sum_{\pi \in \Pi^n} \sum_{m=1}^{n-1}
\underbrace{  \pa_\m \cj_m^\m(x_{\p(1)}, \ldots, x_{\p(m)})
\, N_{n-m}\d(x_{\p(k+1)}, \ldots, x_{\p(n)}) }_{relevant\, contractions}
\ee
where it is understood that only
relevant contractions are
made on the right-hand side.
The factors $N_{n-m}$ are tree-level normalization terms of
the $T$-products~\footnote{When one refers
to local normalization terms  in tree graph contributions in BSEG they
are  always defined with respect to
the natural splitting (see footnote \ref{nat}).}
solution
that contain $(n-m)$ $T_1$ vertices. In all
equations, Wick-ordering is always understood.

\medskip

Now we  have to identify the field-equation terms
$R^{A;m}(\hbar) \ck_{AB} \f^B \d^{(m)}$ in (\ref{offshell}).
Let us recall that, generically,  after natural splitting
(this refers to tree-level graphs, for loop graphs one uses some
reference-splitting solution) we end up with
\be
\pa_\m \cj_{c,m}^{\m} = A_{c,m}.
\ee
The assumption that the quantum Noether method works successfully
means that the ({\it on-shell}) anomaly $A_{c,m}$
is a divergence
up to terms $B_m$ which vanish when the free-field equations are used,
i.e. \be \label{an1}
A_{c,m} = \pa_\m A_{c,m}^\m + * B_m * \, ,
\ee
where $A_{m,c}^\m$ and $B_m$ are some local distributions (since $A_{c,m}$
is local). This decomposition is not unique since one can move derivatives
of field-equation terms from $B_m$ to $A_{c, m}^\m${}.
 This freedom is fixed
 by demanding that $B_m$ contain no  derivatives of field equations
(for details see \cite{paper1}, section 4.2);
thus, we can identify $B_m$ with the right-hand side  of (\ref{offshell}), while
the divergence term $\pa_\m A_{c,m}^\m$ can be canceled  by the
introduction of an additional  normalization term of the $T$-product
$\cj^\m_{n/l}$.

\medskip

An additional ambiguity is related to the additional
global symmetries of the free action. If we  make the transformation
\be \label{amb}
A_{c, m}^\m \to A_{c, m}^\m + \tilde{j^\m} \delta^m; \quad
R^{A;m} \to R^{A;m} - \tilde{\gamma} \f^A \, ,
\ee
where $\tilde{j^\m}$ is a Noether current that generates the
additional symmetry
transformations $\tilde{\gamma} \f^A$, then the right-hand side
of (\ref{an1}) remains unchanged. The meaning of this ambiguity 
is that different 
symmetries can mix with each other. In general, one {\it must}
include in the construction all currents that have non-trivial
commutation relations among themselves. In the case of BRST
symmetry, we have seen in the main text that one should
include all non-trivial global currents of ghost number 1
in the construction
in order to correctly account for all cohomology.

\medskip

Having identified the field-equation terms as the only source
of local terms out of tree contractions
and having clarified the ambiguities due to the field equation
 we  can use the off-shell trick to analyse the quantum Noether condition
 --- first at tree level.
Let us {\it define}
\be
\label{delta}
\gamma_{(m-1)}\f^A = {1 \over  m!} R^{A;m}(\hbar^0); \ \ m > 1.
\ee
We have explicitly shown in \cite{paper1,proc} that,
\be
\gamma \f^A = \sum_{m=0}^k g^m \gamma_m \f^A,
\ee
are symmetry transformation rules that leave the Lagrangian invariant
(up to total derivatives):
\be \label{lagr}
\cl = \sum_{m=0}^{k'} g^m \cl_m,
\ee
where $k$ and $k'$ are integers (which may be infinity).
The Lagrangian $\cl$ is determined from the tree-level normalization
conditions as follows,
\be \label{lagdef}
\cl_m = {\hbar \over i} {N_m \over m!}, \quad {\rm for} \quad m>1,
\ee
where $N_m$ denotes the local normalization ambiguity of
$T_m[T_1(x_1)...T_1(x_m)]$ in tree graphs defined with respect
to the natural splitting  solution. For $m=1$, $\cl_1=(\hbar/i)T_1$.
Note that we regard
(\ref{lagdef}) as the definition of $\cl_m$ within the BSEG approach.
Details can be found in \cite{paper1}
where we also discuss the QNC at loop level.

\end{document}